\documentclass[12pt]{article}
\usepackage{latexsym}
\usepackage{tikz}
\usepackage{amsfonts}
\usepackage{tkz-graph}
\usepackage{amssymb}
\usepackage{comment}
\usepackage{tikz}
\newcommand*\circled[1]{\tikz[baseline=(char.base)]{
            \node[shape=circle,draw,inner sep=2pt] (char) {#1};}}
\usetikzlibrary{shapes}
\usetikzlibrary{snakes}
\usetikzlibrary{shapes}
\usetikzlibrary{arrows}
\usetikzlibrary{decorations.pathmorphing}

\newtheorem{Theorem}{Theorem}[section]

\newtheorem{Lemma}[Theorem]{Lemma}
\newtheorem{Observation}[Theorem]{Observation}

\newtheorem{Problem}{Problem}[section]

\newcommand{\qed}{\hfill $\Box$}

\def\inst#1{$^{#1}$}
\def\CCD{($4K_1$, $C_4$, $C_6$)}

\title {The intersection of two vertex coloring problems}

\author{
  Ang\`ele M. Foley\inst{1} \footnote{Formerly Ang\`ele M. Hamel}
    \and Dallas J. Fraser\inst{1}
    \and Ch\'inh T. Ho\`ang\inst{1}
    \and Kevin Holmes \inst{1}
    \and Tom P. LaMantia \inst{1} 
}
\begin{document}
\maketitle

\begin{center}
{\footnotesize

\inst{1} Department of Physics and Computer Science, Wilfrid Laurier
University, \\Waterloo, Ontario, Canada
}

\end{center}

\begin{abstract}
A {\it hole} is an induced cycle with at least four vertices. A hole is even if its number of vertices is even. Given a set $L$ of graphs, a graph $G$ is $L$-free if $G$ does not
contain any graph in $L$ as an induced subgraph. Currently, the following two problems
are unresolved: the complexity of coloring even hole-free graphs, and the complexity of coloring $\{4K_1, C_4\}$-free graphs. The intersection of these two problems is the problem of coloring $\{4K_1, C_4, C_6\}$-free graphs. In this paper we present partial results on this problem. 

\end{abstract}

\section{Introduction}
There has been recently keen interest in finding polynomial-time algorithms to optimally color graphs $G$ that do not contain any graph in a list $L$ as an induced subgraphs. 
Particular attention is focused on graphs  whose forbidden list $L$
contains graphs with four vertices, and recent papers of Lozin
and Malyshev \cite{Lozin}, and of Fraser, Hamel, Ho\`ang, Holmes and LaMantia \cite{FraHam2017}  discuss the state of the art on this
problem,  identifying three outstanding classes: $L=(4K_1$, claw),
$L=(4K_1$, claw, co-diamond), and $L=(4K_1, C_4$).  As a
resolution of these cases is likely challenging, a productive
approach would be to consider increasing the number of graphs in $L$. But it makes
sense to ask, which $L$ is an interesting one to consider? We are particularly interested in cases that are slightly larger than the class $L=(4K_1, C_4$), which is one of the unresolved cases. Before we introduce the problems we need some definitions:

%
A {\it hole} is an induced cycle with at least four vertices. A hole is even if its number of vertices is even. The problem of coloring even-hole-free graphs has been much studied. A theorem of Addario-Berry, Chudnovsky, Havet, Reed, and 
Seymour \cite{AddChu2008} shows that for an even-hole-free graph $G$, the chromatic number of $G$ is at most two times its clique number (the number of vertices in a largest clique of $G$). It is currently not known whether even-hole-free graphs can be colored in polynomial time.

We offer the following four problems for consideration:
\begin{Problem}\label{pro:main}
What is the complexity of coloring ($4K_1, C_4$)-free graphs?
\end{Problem}
This is the original problem, and is likely the most challenging.
\begin{Problem}\label{pro:even-hole}
What is the complexity of coloring even-hole-free graphs?
\end{Problem}
Problem~\ref{pro:even-hole} might even be NP-complete.  Combining Problems~\ref{pro:main} and \ref{pro:even-hole}, we get the following problem:
\begin{Problem}\label{pro:4k1-even-hole}
	What is the complexity of coloring ($4K_1$, even hole)-free graphs?
\end{Problem}
This problem appears to be more tractable than the previous two. In this paper, we study Problem~\ref{pro:4k1-even-hole}. Since a $4K_1$-free graph does not contain a hole of length at least 8, Problem~\ref{pro:4k1-even-hole} is equivalent to the following:
\begin{Problem}\label{pro:c6}
	What is the complexity of coloring ($4K_1, C_4, C_6$)-free graphs?
\end{Problem}
Even though we have not been able to solve Problem~\ref{pro:c6},  we have succeeded, in some sense, in solving ``half'' of it, as follows.
Consider a ($4K_1, C_4, C_6$)-free graph $G$. We may assume $G$ is not perfect (there are known algorithms to color perfect graphs). Thus $G$ has to contain a $C_5$ or $C_7$. If $G$ contains a $C_7$, then our result shows that $G$ can be colored in polynomial time. The case where $G$ contains a $C_5$ but not a $C_7$ is open. Investigation into this problem led us to a proof that there is a polynomial time algorithm to color a ($4K_1, C_4, C_6$, $C_5$-twin)-free graph. 

In Section~\ref{sec:background}, we discuss the background of the problem and state the main results. In Section~\ref{sec:c7}, we study ($4K_1, C_4, C_6$)-free graphs that contain a $C_7$ and show that such graphs can be colored in polynomial time. In Section \ref{sec:c5}, we study ($4K_1, C_4, C_6$)-free graphs that contain a $C_5$, but no $C_7$. In Section \ref{sec:clique-cutset}, we give a polynomial time algorithm to color a ($4K_1, C_4, C_6$, $C_5$-twin)-free graph. Finally, in Section \ref{sec:conclusion}, we dicuss open problems related to our work.

\section{Background and results}\label{sec:background}
 
Before discussing our results in more detail, we need introduce a
few definitions.
Let $G$ be a graph.  A {\em colouring} of a graph $G=(V,E)$ is  a
mapping $f: V \rightarrow \{ 1, \ldots, k\}$  for some nonnegative
integer $k$ such that $f(u) \neq f(v)$ whenever $uv\in E$. The
{\em chromatic number}, denoted $\chi(G)$, is the minimum number
of colors needed to colour a graph $G$. {\em VERTEX COLORING} is
the problem of determining the chromatic number of a graph.

Consider the following operations to build a graph.
\begin{description}
\item [(i)] Create a vertex $u$ labeled by integer $\ell$.

\item [(ii)] Disjoint union (i.e., co-join)

\item [(iii)] Join between all vertices with label $i$ and all
vertices with label $j$ for $i \not=
 j$, denoted by $\eta_{i,j}$ (that is, add all edges between vertices of label $i$ and label $j$).

\item [(iv)] Relabeling all vertices of label $i$ by label $j$,
denoted by $\rho_{i \rightarrow j}$
\end{description}
The {\it clique width} of a graph $G$, denoted by $cwd(G)$, is the
minimum number of labels needed to build the graph with the above
four operations. It is well-known \cite{CouOla2000} that if the
clique width of a graph is bounded then so is that of its
complement. Clique widths have been intensively studied. In Rao
\cite{Rao}, the following result is established.
\begin{Theorem}\label{thm:Rao2007}
VERTEX COLORING is polynomial time solvable for graphs with
bounded clique width. \qed
\end{Theorem}
We will need the folowing well known observation that is easy to establish (for example, see \cite{Brandstadt}).
\begin{Observation}\label{obs:folklore}
	Let $G$ be a graph and $G'$ be the graph obtained from $G$ by removing a constant 
	number of vertices. Then $G$ has bounded clique width if and only if $G'$ does. \qed
\end{Observation}

The symbol $\omega(G)$ denotes the number of vertices in a largest
clique of $G$.   A graph $G$ is {\it perfect} if for each induced
subgraph $H$ of $G$, we have $\chi(H) = \omega(H)$. A {\em hole}
is an induced cycle of length at least $4$, i.e. $C_k$ for $k\geq
4$. A hole is {\em even} or {\em odd} depending on the parity of
the vertices in the hole.  An {\em anti-hole} is the complement of
a hole.

Two important results are known about perfect graphs.  The Perfect
Graph Theorem, proved by Lov\'asz \cite{Lov1972}, states that a
graph is perfect if and only if its complement is. The Strong
Perfect Graph Theorem, proved by Chudnovsky,  Robertson, Seymour,
and  Thomas \cite{ChuRob2006}, states that a graph is perfect if
and only if it is odd-hole-free and odd-anti-hole-free. Both of
the above results were long standing open problems proposed by
Berge \cite{Ber1961}. Gr\"otschel, Lov\'asz and Schrijver
\cite{GroLov1984} designed a polynomial-time  algorithm for
finding a largest clique and a minimum coloring of a perfect
graph.

\begin{figure}
	\begin{center}
		\begin{tikzpicture} [scale = 1.25]
		\tikzstyle{every node}=[font=\small]
		
		\newcommand{\size}{1}

		\newcommand{\cfivetwin}{4}{
			\path (\size * 6, 0) coordinate (g4);
			\path (g4) +(0, 0) node (g4_1){};
			\path (g4) +(0, \size) node (g4_2){};
			\path (g4) +(\size * 0.5, \size*1.25) node (g4_3){};
			\path (g4) +(\size, \size) node (g4_4){};
			\path (g4) +(\size, 0) node (g4_5){};
			\path (g4) +(\size * 0.5, \size * 0.75) node (g4_6){};
			\foreach \Point in {(g4_1), (g4_2), (g4_3), (g4_4),(g4_5),(g4_6)}{
				\node at \Point{\textbullet};
			}
			\draw   (g4_1) -- (g4_2)
			(g4_1) -- (g4_5)
			(g4_2) -- (g4_3)
			(g4_2) -- (g4_6)
			(g4_3) -- (g4_4)
			(g4_3) -- (g4_6)
			(g4_4) -- (g4_5)
			(g4_4) -- (g4_6);
			
			\path (g4) ++(\size / 2, -\size / 2) node[draw=none,fill=none] { {\large $C_5-Twin$}};
		}

		\end{tikzpicture}
	\end{center}
	\caption{The $C_5$-twin}\label{fig:c5-twin}
\end{figure}
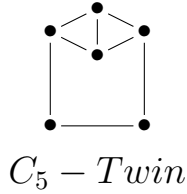

Suppose we want to color a ($4K_1, C_4, C_6$)-free graph $G$. Note that $G$ contains no $C_\ell$ for $\ell \geq 8$ because $G$ is $4K_1$-free. By
the result of Gr\"otschel et al, we may assume $G$ is not perfect.
The result of Chudnovsky et al implies $G$ contains a $C_5$ or $C_7$ as an
induced subgraph (note that the anti-hole of length at least six
contains a $C_4$ and the $C_5$ is self-complementary.) If $G$ contains a $C_7$, 
then we will show that $G$ has bounded clique-width. 
\begin{Theorem}\label{thm:main-c7}
Let $G$ be ($4K_1, C_4, C_6$)-free graph that contains a $C_7$.
Then $G$ has bounded clique width. 
\end{Theorem}
We will use Theorem~\ref{thm:main-c7} to prove the following theorem that is the main result of this paper.
\begin{Theorem}\label{thm:main-coloring}
	VERTEX COLORING can be solved in polynomial time for the class of ($4K_1, C_4, C_6$)-free graphs that contain a $C_7$.
\end{Theorem}
Two adjacent vertices $x,y$ of a graph $G$ are {\it twins} if for any vertex $z$ different from $x$ and $y$, $xz$ is an edge if and only if $yz$ is an edge.  A {\it hole-twin} is the graph obtained from a hole by adding a vertex that form twins with some vertex of the hole. Figure~\ref{fig:c5-twin} shows the $C_5$-twin. Hole-twins play an interesting role in graph theory. They are among the forbidden induced subgraphs for line-graphs (Beineke \cite{Bei1970}). 

By adding the $C_5$-twin to the list of forbidden induced subgraphs for our graph class, we obtain the following theorem. 
\begin{Theorem}\label{thm:c5-twin-color}
	VERTEX COLORING can be solved in polynomial time for the class of  ($4K_1, C_4, C_6$, $C_5$-twin)-free graphs.
\end{Theorem}

We will rely on a theorem of Fraser et al \cite{FraHam2017}. To explain this theorem, we will need to introduce a few definitions.  Given sets of
vertices $X,Y$, we write $X \;\circled{0}\; Y$ to mean there is no
edge between any vertex in $X$ and any vertex in $Y$ (also called
a {\em co-join}). Given sets of vertices $X, Y$, we write $X
\;\circled{1}\; Y$ to mean there are all edges between $X$ and $Y$
(also called a {\em join}).

Consider a partition ${\cal P}$ of  the vertices of $G$ into sets
$S_1, S_2, \ldots , S_k$ such that each $S_i$ induces a clique. A set $S_i$ is {\it uniform} to a set
$S_j$ if $S_i$ \circled{0} $S_j$ or $S_i$ \circled{1} $S_j$. The set $S_i$ is {\it uniform in} ${\cal P}$ if every set $S_j , i \not= j$, is uniform to it.   The partition ${\cal P}$ is {\it uniform} if every set $S_i$ is uniform in ${\cal P}$.   
A set $S_i$ is {\it near-uniform} if there is at most one set $S_j, i \not= j,$
that is not uniform to $S_i$.
The partition ${\cal P}$ is {\it
near-uniform} if every set $S_i$ is near-uniform. Thus, a uniform partition is near-uniform.
A {\it
$k$-near-uniform} partition of $G$ is a partition of the vertices
of $G$ into $k$ near-uniform sets. In such a partition, the pair
of sets $S_i$, $S_j$ such that each set is not uniform to the
other is call a {\it uniform-pair}. The following is proved in Fraser et al \cite{FraHam2017}.

\begin{Theorem}\label{thm:cliquewidth}\cite{FraHam2017}
Let $G$ be a graph admitting a $k$-near-uniform  partition\footnote{We note that the definition of near-uniform partition in \cite{FraHam2017} is incomplete. The sets $S_i$'s must be cliques for the theorem to hold.} such
that the uniform pairs are $C_4$-free. Then we have $cwd(G)  \leq
2k $. \qed
\end{Theorem}
We will establish the following theorem.
\begin{Theorem}\label{thm:main-uniform}
Let $G$ be ($4K_1, C_4, C_6$)-free graph that contains a $C_7$.
Then $G - C_7$ admits a $k$-uniform  partition, for some constant $k$. 
\end{Theorem}

By the above discussion, Theorem~\ref{thm:main-uniform} implies Theorem~\ref{thm:main-coloring}. We will prove Theorem~\ref{thm:main-uniform} in the next section.

\section{When the graphs contain a $C_7$}\label{sec:c7}

Assume that the graph $G$
is ($4K_1, C_4, C_6$)-free and contains a $C_7$. 
In this section, we examine the structure of the neighborhood of the $C_7$. Then we will prove Theorem~\ref{thm:main-uniform}. 

We will need first to establish 
a number of preliminary results.
Given a
hole, $H$, and a vertex $x$ not in $H$, we say $x$ is a $k$-vertex
(for $H$) if $x$ has exactly $k$ neighbours in $H$. 

For all the claims below, we shall now assume that $G$ contains an induced $C_7$ with vertices $(1,2,3,4,5,6,7)$. The vertex numbers of the $C_7$ are taken modulo 7.

Let 
\begin{itemize}
	\item $X_i$  denote the set of $3$-vertices adjacent to $(i,i+1,i+2)$ in the $C_7$, 
	\item $Y_i$  denote the set of $3$-vertices on $(i,i+1,i+4)$ in the $C_7$,
	\item $Z_i$  denote the set of $5$-vertices on $(i,i+1,i+2,i+3,i+4)$ in the $C_7$, and 
	\item $W$  denote the set of $7$-vertices in the $C_7$.
\end{itemize}

We will show that the sets $X_i, Y_i, Z_i, W$ form a partition of $V(G) - C_7$.

\begin{Observation}\label{obs:no-4vertex-c7}
	The $C_7$ has no $k$-vertex in $G$ for $k = 0,1,2,4,6$.
\end{Observation}
\noindent {\it Proof}. Suppose that there is a $0$-vertex $v$ for $C_7$. This creates an induced $4K1$ $(v,i,i+2,i+4)$, which is forbidden. Therefore, there is no $0$-vertex. Now suppose that there is a $1$-vertex $v$ that is adjacent to $i$. This creates an induced $4K1$ $(v,i+1,i+3,i+5)$, which is forbidden. Therefore, there is no $1$-vertex. Next, suppose that there is a $2$ vertex $v$ for $C_7$. It is easy to see that $v$ and some three vertices in the $C_7$ form a $4K_1$, a contradiction. Consequently, there is no $2$-vertex. Next, suppose that there is a $4$-vertex $v$ for $C_7$. Then $G$  contains a $C_4$ or $C_6$, both of which are forbidden. Therefore, there is no $4$-vertex. Finally, suppose that there exists a $6$-vertex $v$ for $C_7$. Then, $G$ contains an induced $C_4$, which is forbidden. Therefore, there is no $6$-vertex for $C_7$. \qed

It is a routine matter to verify the two observations below. 
\begin{Observation}\label{obs:3-vertex-c7}
	Let $v$ be a 3-vertex for the $C_7$. Then $v \in X_i \cup Y_i$ for some $i$. \qed
\end{Observation}

\begin{Observation}\label{obs:5-vertex-c7}
	Let $v$ be a 5-vertex for the $C_7$. Then $v \in Z_i $ for some $i$. \qed
\end{Observation}

The above three observations imply the following observation.
\begin{Observation}\label{obs:partition}
	Let $G'$ be the graph obtained from $G$ by removing the $C_7$. Then the sets $X_i, Y_i, Z_i, W$ form a partition ${\cal P}$ of $G'$.
\end{Observation}

 Our aim is to show that $G'$ has bounded clique width. In fact, we will show that ${\cal P}$ is an uniform partition of $G'$. 

We now examine the adjacencies between the sets of the partition ${\cal P}$. 
\begin{Observation}\label{obs:each-set-is-clique}
	Each of the sets $X_i, Y_i, Z_i, W$ of the partition ${\cal P}$ is a clique.
\end{Observation}
{\it Proof}. It is easy to see that if a set of ${\cal P}$ is not a clique then there is a $C_4$. \qed

The next sequence of observations will imply that $X_i$ is near-uniform in ${\cal P}$.
	
\begin{Observation}\label{obs:Xi-begin}
		$X_i$ \circled{1} $X_{i+1} \cup X_{i+6} $
\end{Observation}
\noindent {\it Proof}.  Suppose there are vertices $x_1 \in X_i$ and $x_2 \in X_{i+1}$ such that $x_1x_2 \notin E$. Then there exists a $4K_1$ $(x_1,x_2,i+4,i+6)$, which is forbidden. Therefore, $X_i$ \circled{1} $X_{i+1}$. By symmetry, we have $X_i$ \circled{1} $X_{i+6}$.  \qed
	
\begin{Observation}
	$X_i$ \circled{0} $X_{i+2} \cup X_{i+3} \cup X_{i+4} \cup X_{i+5}$.
\end{Observation}
\noindent {\it Proof}. Consider a vertex $x_1 \in X_i$. 
Suppose there is a vertex $x_2 \in X_{i+2}$ such that $x_1x_2 \in E$. This creates a $C_6$ $(x_1, x_2, i+4,i+5,i+6,i)$, which is forbidden. Therefore, we have $X_i$ \circled{0} $X_{i+2}$, and by symmetry $X_i$ \circled{0} $X_{i+5}$. Now suppose that there is a vertex $x_3 \in X_{i+3}$ such that $x_1x_3 \in E$. This creates a $C_4$ $(x_3,x_1,i+2,i+3)$, which is forbidden. Therefore, we have  $X_i$ \circled{0} $X_{i+3}$, and by symmetry, $X_i$ \circled{0} $X_{i+4}$. \qed
	
\begin{Observation}
	$X_i$ \circled{1} $Y_i \cup Y_{i+1} \cup Y_{i+4}$.
\end{Observation}
\noindent{\it Proof}. Consider a vertex $x \in X_i$. Suppose there is a vertex $y \in Y_i$ such that $xy \notin E$. This creates a $4K1$ $(x,y,i+3,i+5)$, which is forbidden. Therefore, $X_i$ \circled{1} $Y_i$. Now suppose there is a vertex $y \in Y_{i+1}$ with $xy \notin E$. This creates a $4K_1$ $(x,y,i+3,i+6)$, which is forbidden. Therefore $X_i$ \circled{1} $Y_{i+1}$. Finally, suppose there is a vertex $y \in Y_{i+4}$ with $xy \notin E$. This creates a $4K_1$ $(x,y,i+3,i+6)$, which is forbidden. Therefore $X_i$ \circled{1} $Y_{i+4}$. Consequently, $X_i$ \circled{1} $Y_i \cup Y_{i+1} \cup Y_{i+4}$.  \qed
	
\begin{Observation}
		$X_i$ \circled{0} $Y_{i+2} \cup Y_{i+3} \cup Y_{i+5} \cup Y_{i+6}$
\end{Observation}
\noindent{\it Proof}. Consider a vertex $x \in X_i$. Suppose there is a vertex $y \in Y_{i+2}$ with $xy \in E$. This creates a $C_4$ $(x,y,i+6,i)$, which is forbidden. Therefore, we have $X_i$ \circled{0} $Y_{i+2}$, and by symmetry, $X_i$ \circled{0} $Y_{i+6}$. Now suppose that there is a vertex $y \in Y_{i+3}$ with $xy \in E$. This creates a $C_4$ $(x,y,i+3,i+2)$, which is forbidden. Therefore, $X_i$ \circled{0} $Y_{i+3}$, and by symmetry, $X_i$ \circled{0} $Y_{i+5}$. \qed

	
\begin{Observation}
		$X_i$ \circled{1} $Z_i \cup Z_{i+1} \cup Z_{i+4} \cup Z_{i+5} \cup Z_{i+6}$
\end{Observation}
\noindent{\it Proof}. Consider a vertex $x \in X_i$. Suppose there is a vertex $z \in Z_{i}$ such that $xz \notin E$. This creates a $C_4$ $(x,i,z,i+2)$, which is forbidden. Therefore, we have $X_i$ \circled{1} $Z_i$, and by symmetry, $X_i$ \circled{1} $Z_{i+5}$. 
Now suppose there is a vertex $z \in Z_{i+1}$ such that $xz \notin E$. This creates a $C_6$ $(i,i+6,i+5,z,i+2,x)$, which is forbidden. Therefore, we have $X_i$ \circled{1} $Z_{i+1}$, and by symmetry, $X_i$ \circled{1} $Z_{i+4}$. 
Finally, suppose that there is a vertex $z \in Z_{i+6}$ such that $xz \notin E$. This creates a $C_4$ $(x,i,z,i+2)$, which is forbidden. Therefore, we have $X_i$ \circled{1} $Z_{i+6}$, and we are done. \qed
	
\begin{Observation}
	$X_i$ \circled{0} $Z_{i+2} \cup Z_{i+3}$
\end{Observation}
\noindent{\it Proof}. Consider a vertex $x \in X_i$. Suppose there is a vertex $z \in Z_{i+2}$ such that $xz \in E$. This creates a $C_4$ $(x,z,i+6,i)$, which is forbidden. Therefore, $X_i$ \circled{0} $Z_{i+2}$. Next, suppose that there is a vertex $z \in Z_{i+3}$ such that $xz \in E$. This also creates a $C_4$ $(x,z,i+3,i+2)$, which  is forbidden. Therefore, $X_i$ \circled{0} $Z_{i+3}$. \qed
	
\begin{Observation}\label{obs:Xi-end}
	$X_i$ \circled{1} $W$
\end{Observation}
\noindent{\it Proof}. Suppose that there are vertices $x \in X_i$ and $w \in W$ such that  $xw \notin E$. This creates a $C_4$ $(i,x,i+2,w)$, which is forbidden. Therefore we have $X_i$ \circled{1} $W$. \qed

Observations~\ref{obs:Xi-begin}--\ref{obs:Xi-end} together imply the following lemma.
\begin{Lemma}\label{lem:Xi-uniform}
 	For every $i$, the set $X_i$ is uniform in the partition ${\cal P}$. \qed
\end{Lemma}
Next, we examine the sets $Y_i$.
\begin{Observation}\label{obs:at-most-two-distinct-y}
	At most two of the sets $Y_1, Y_2, \ldots , Y_7$ can be non-empty. In particular, if $Y_i \not= \emptyset$, then $Y_{i+1} = Y_{i+2} = Y_{i+5} = Y_{i+6} = \emptyset$, and we have $Y_{i+3} \neq \emptyset$ or $Y_{i+4} \neq \emptyset$, but not both.
\end{Observation}
\noindent {\it Proof}. Suppose $Y_i \not= \emptyset$. Let  $y$ be a vertex in $ Y_i$. Suppose the set $Y_{i+1}$ contains a vertex  $y_1$. If $y y_1 \notin E$, then there exists a $4K_1$ $(y,y_1,i+3,i+6)$, which is forbidden. If $y y_1 \in E$, then there is a $C_4$ ($y, y_1, i+5, i+4$), which is forbidden. So we have $Y_{i+1} = \emptyset$, and by symmetry, $Y_{i+6} = \emptyset$. Now, suppose $Y_{i+2}$ is non-empty and contains a vertex $y_2$. If $y y_2 \in E$, then there is a $C_4$ ($y, i+4, i+3, y_2$), a contradiction. But if $y y_2 \notin E$, then there exists a $C_6$ $(y,i+4,i+3,y_2,i+6,i)$, a contradiction. So, we have $Y_{i+2} = \emptyset$, and by symmetry, $Y_{i+5} = \emptyset$. 

The first part of this proof shows that if $Y_{i+3} \not= \emptyset$, then $Y_{i+4} = \emptyset$. So only one of the two sets $Y_{i+3}, Y_{i+4}$ can be non-empty. \qed

\begin{Observation}
	$Y_i$ \circled{1}  $Y_{i+3} \cup Y_{i+4}$ 
\end{Observation}
\noindent{\it Proof}. Consider a vertex $y \in Y_i$. Suppose there is a vertex $y_3 \in Y_{i+3}$ such that $yy_3 \notin E$. There is a $C_4$ $(y,i,y_3,i+4)$, a contradiction. So, we have $Y_i$ \circled{1}  $Y_{i+3}$, and by symmetry, $Y_i$ \circled{1}  $Y_{i+4}$. \qed

\begin{Observation}\label{obs:y-restricts-z-c7}
	If $Y_i \neq \emptyset$, then $Z_{i+5} = Z_{i+6} = \emptyset$
\end{Observation}
\noindent{\it Proof}. Assume that $Y_i \neq \emptyset$ and $Z_{i+5} \neq \emptyset$. Consiser vertices   $y \in Y_i$ and $z \in Z_{i+5}$. If $yz \in E$, then there exists a $C_4$ $(y,z,i+5,i+4)$, which is forbidden. However, if $yz \notin E$, then there exists a $C_6$ $(y,i,z,i+2,i+3,i+4)$, which is also forbidden. Therefore if $Y_i \neq \emptyset$, then $Z_{i+5} = \emptyset$, and by symmetry, $Z_{i+6} = \emptyset$. \qed
	
\begin{Observation}
	$Y_i$ \circled{1} $W \cup Z_i \cup Z_{i+1} \cup Z_{i+3} \cup Z_{i+4}$
\end{Observation}
\noindent{\it Proof}. Consider a vertex $y \in Y_i$. Let $w \in W$. If $yw \notin E$, then there is a $C_4$ $(i,w,i+4,y)$, which is forbidden. Therefore, $Y_i$ \circled{1} $W$. 
 
Consider a vertex $z \in Z_i \cup Z_{i+1} \cup Z_{i+3} \cup Z_{i+4}$. Then $z (i+4) \in E$, and $z$ is adjacent to a vertex $i' \in \{i, i+1\}$. If $zy \notin E$, there there is a $C_4$ $(i',y,i+4,z)$, which is forbidden. Therefore,  $Y_i$ \circled{1} $Z_i \cup Z_{i+1} \cup Z_{i+3} \cup Z_{i+4}$. \qed

\begin{Observation}\label{obs:Yi-end}
		$Y_i$ \circled{0} $Z_{i+2}$
\end{Observation}
\noindent{\it Proof}.Consider vertices $y \in Y_i$ and $z \in Z_{i+2}$ such that $yz \in E$. This creates a $C_4$ $(y,z,i+6,i)$, which is forbidden. Therefore $Y_i$ \circled{0} $Z_{i+2}$. \qed

Observations~\ref{obs:at-most-two-distinct-y}--\ref{obs:Yi-end} together imply the following lemma. 	
\begin{Lemma}\label{lem:Yi-uniform}
	For every $i$, the set $Y_i$ is uniform in the partition ${\cal P}$. \qed
\end{Lemma}
Next, we examine the sets $Z_i$.
\begin{Observation}\label{obs:z-restrictions-c7}
	If $Z_i \neq \emptyset$, then $Z_{i+2} = \emptyset$ and $Z_{i+5} = \emptyset$.
\end{Observation}
\noindent{\it Proof}. Suppose that $Z_i \neq \emptyset$.  Also, suppose  $Z_{i+2} \neq \emptyset$. Consider  vertices $z_i \in Z_i$ and $z_{i+2} \in Z_{i+2}$. If $z_i z_{i+2} \in E$, then there is a $C_4$ ($z_i, z_{i+2}, i+6, i$), a contradiction. If $z_i z_{i+2} \not\in E$, then there is a $C_4$ ($z_i, i+4, z_{i+2}, i+2$), a contradiction. So $Z_{i+2}$ is empty, and by symmetry, $Z_{i+5}$ is empty. \qed

\begin{Observation}
	There can exist at most $3$ distinct sets of $5$-vertices for $C_7$.
\end{Observation}
\noindent{\it Proof}. Follows from Observation \ref{obs:z-restrictions-c7}.
	
\begin{Observation}\label{obs:z-join}
		$Z_i$ \circled{1} $W  \cup Z_{i+1}  \cup Z_{i+3}  \cup Z_{i+4}  \cup Z_{i+6}$
\end{Observation}
\noindent{\it Proof}. Consider a vertex $z_i \in Z_i$ and a vertex $z \in V(G) - C_7$ with $z \not= z_i$. If $z$ is adjacent to two non-adjacent vertices, say $a$ and $b$, of the set $\{i, i+1, i+2, i+3, i+4\}$, then $z z_i \in E$, for otherwise there is a $C_4$ ($z, a, z_i, b$). Observe that any vertex in $W  \cup Z_{i+1}  \cup Z_{i+3}  \cup Z_{i+4}  \cup Z_{i+6}$ is adjacent to two non-adjacent vertices of $\{i, i+1, i+2, i+3, i+4\}$. The Observation follows. \qed

Observations~\ref{obs:z-restrictions-c7}--\ref{obs:z-join} together imply the following lemma. 	
\begin{Lemma}\label{lem:Zi-uniform}
	For every $i$, the set $Z_i$ is uniform in the partition ${\cal P}$. \qed
\end{Lemma}
We can now prove our main results. 

\noindent {\it Proof of Theorem~\ref{thm:main-uniform}}. Let $G$ be a $(4K_1, C_4, C_6$)-free graph with a $C_7$. Define the sets $X_i, Y_i, Z_i, W$ as above. Let $G' = G - C_7$. Observations~\ref{obs:partition} implies that the sets $X_i, Y_i, Z_i, W$ form a partition  of $G'$. Lemmas~\ref{lem:Xi-uniform}, \ref{lem:Yi-uniform}, and \ref{lem:Zi-uniform} show that each of the sets $X_i, Y_i, Z_i$ is uniform. Now the set $W$ is also uniform because every other set is uniform to $W$. Thus, the partition is uniform. \qed

\noindent {\it Proof of Theorem~\ref{thm:main-c7}}.  Let $G$ be a $(4K_1, C_4, C_6$)-free graph with a $C_7$. The graph $G' = G - C_7$ has bounded clique width by Theorem~\ref{thm:main-uniform} and  Theorem~\ref{thm:cliquewidth}, $G'$ has bounded clique width. Thus, $G$ has bounded clique width by Observation~\ref{obs:folklore}. \qed

\noindent {\it Proof of Theorem~\ref{thm:main-coloring}}. Let $G$ be a $(4K_1$, even hole)-free graph with a $C_7$. By Theorem~\ref{thm:main-c7}, $G$ has bounded clique width. By Theorem~\ref{thm:Rao2007}, $G$ can be optimally colored in polynomial time. \qed

So, if our graphs contain a $C_7$, we know how to color them. If they do not contain a $C_7$, then we know they must contain a $C_5$, for otherwise they are perfect and we would know how to color them. In the next section, we discuss the case of the $C_5$.

\section{When the graphs contains a $C_5$}\label{sec:c5}

In this section, we assume $G$ is {\CCD}-free. For the all the claims below, we will also assume that $G$ contains an induced $C_5$ with vertices $(i,i+1,i+2,i+3,i+4)$. Let $R$ denote the set of $0$-vertices for this $C_5$, let $F_i$ be the set of $1$-vertices adjacent to $i$, let $T_i$ be the set of $2$-vertices with neighbors $(i,i+1)$, let $X_i$ be the set of $3$-vertices with neighbors $(i,i+1,i+2)$ and let $W$ denote the set of $5$-vertices.

The following observation is immediate.	
\begin{Observation}\label{obs:partition-for-c5}
	The sets $F_i, T_i, X_i, R, W$ form a partition of the vertex set of $G-C_5$ \qed
\end{Observation}
\begin{Observation}\label{obs:r-clique}
	Each of $F_i, T_i, X_i, R, W$ form a clique.
\end{Observation}
\noindent {\it Proof}. Consider two non-adjacent vertices $x,y$ of $G$. If both $x,y$ belong to $F_i$, then $x$ and $y$ and some two non-adjacent vertices of the $C_5$ form a $4K_1$, a contradiction. Similarly, we can see that $x,y$ cannot both belong to $T_i$, or to $R$. If $x,y$ both belong to $X_i$ or to $W$, then $x,y$ and some two non-adjacent vertices of $C_5$ form a $C_4$. \qed
\begin{Observation} \label{obs:r-join-Fi-union-Ti}
	$R$ \circled{1} $F_i \cup T_i$. 
\end{Observation}
\noindent {\it Proof}.  Consider a vertex $r \in R$ and a vertex $s \in F_i \cup T_i$. If $rs \notin E$, then $r,s$ and some two non-adjacent vertices in the $C_5$ form a $4K_1$. \qed

\begin{Observation}\label{obs:one-Fi}
	If $F_i \neq \emptyset$ then $F_j = \emptyset$ for all $j \neq i$.
\end{Observation}
\noindent {\it Proof}. Consider vertices $f_i \in F_i$,  $f_j \in F_j$, with $i \not= j$. We must have $f_i f_j \in E$, for otherwise $f_i,f_j$ and some two vertices in the $C_5$ form a $4K_1$. If $j = i+1$, then there is a $C_4$ ($f_i, f_j, j, i $). So we have $j \not = i+1$, and by symmetry, $j \not= i-1$. If $j = i+2$, then there is a $C_6$ ($f_i, f_j, j, j+1, j+2, i $). So we have $j \not = i+2$, and by symmetry, $j \not= i-2$. \qed
\begin{Observation} \label{obs:F-join-Ti}
	$F_i$ \circled{1} $T_i \cup T_{i+2} \cup T_{i+4}$. 
\end{Observation}
\noindent {\it Proof}. Let $f \in F_i$, and let $t \in T_i \cup T_{i+2}$. If $f t_i \notin E$, then $f,t$ and some two non-adjacent vertices of the $C_5$ form a $4K_1$. So $F_i$ \circled{1} $T_i \cup T_{i+2}$. By symmetry (with the case $T_i$), we have $F_i$ \circled{1} $T_{i+4}$. \qed

\begin{Observation} \label{obs:F-cojoin-Ti}
	$F_i$ \circled{0} $T_{i+1} \cup T_{i+3}$. 
\end{Observation}
\noindent {\it Proof}. Let $f \in F_i$, and let $t \in T_{i+1}$. If $ft \in E$, then there is a $C_4$ ($f,t, i+1, i$). So we have $F_i$ \circled{0} $T_{i+1}$, and by symmetry, $F_i$ \circled{0} $T_{i+3}$. \qed

\begin{Observation} \label{obs:F-cojoin-Xi+1}
	$F_i$ \circled{0} $X_{i+1}$. 
\end{Observation}
\noindent {\it Proof}. Let $f \in F_i$, and let $x \in X_{i+1}$. If $fx \in E$, then there is a $C_4$ ($f,t, i+1, i$). So we have $F_i$ \circled{0} $X_{i+1}$. \qed

We note that vertices of $F_i$ may have neighbors and non-neighbors in $X_i \cup X_{i+3} \cup X_{i+4}$.

\begin{Observation} \label{obs:Ti-cojoin-Tj}
	$T_i$ \circled{0} $T_j$ for all $j \not = i$. 
\end{Observation}
\noindent {\it Proof}. Consider vertices $t_i \in T_i, t_{i+1} \in T_{i+1}$. If $t_i t_{i+1} \in E$, then there is a $C_6$ ($t_i, t_{i+1}, i+2, i+3, i+ 4, i$). So we have $T_i$ \circled{0} $T_{i+1}$, and by symmetry, $T_i$ \circled{0} $T_{i+4}$. Now consider a vertex  $t_{i+2} \in T_{i+2}$. If $t_i t_{i+2} \in E$, then there is a $C_4$ ($t_i, i+1, i+2, t_{i+2}$). So we have $T_i$ \circled{0} $T_{i+2}$, and by symmetry, $T_i$ \circled{0} $T_{i+3}$. \qed

\begin{Observation} \label{obs:Ti-cojoin-Xi+2}
	$T_i$ \circled{0} $X_{i+2}$. 
\end{Observation}
\noindent {\it Proof}. Consider vertices $t_i \in T_i, x_{i+2} \in X_{i+2}$. If $t_i x_{i+2} \in E$, then there is a $C_4$ ($t_i, x_{i+2}, i+2, i+1$). \qed

\begin{Observation} \label{obs:Xi-cojoin-Xi+2}
	$X_i$ \circled{0} $X_{i+2}$. 
\end{Observation}
\noindent {\it Proof}. Consider vertices $x_i \in X_i, x_{i+2} \in X_{i+2}$. If $x_i x_{i+2} \in E$, then there is a $C_4$ ($x_i, x_{i+2}, i+4, i$). \qed

In the next section, we will use the results of this section to prove Theorem~\ref{thm:c5-twin-color}.

\section{Clique cutset decomposition}\label{sec:clique-cutset}
In this section, we present a proof of Theorem~\ref{thm:c5-twin-color}. We will need to introduce
definitions and background for the problem.

Consider a graph $G$.
A {\em clique cutset} of $G$ is a set of vertices $S$ such that
$S$ is a clique and $G-S$ is disconnected.  
Consider the following procedure to
decompose $G$. If $G$ has a clique cutset $C$, then $G$ is 
decomposed into subgraphs $G_1 = G[V_1]$ and $G_2 = G[V_2]$ 
where $V = V_1 \cup V_2$ and $C = V_1 \cap V_2$ ($G[X]$ denotes 
the subgraph of $G$ induced by $X$ for a subset $X$ of vertices 
of $V(G)$). Given optimal colourings of $G_1, G_2$, we can 
obtain an optimal colouring of $G$ by identifying the colouring 
of $C$ in $G_1$ with that of $C$ in $G_2$. In particular, we 
have $\chi(G)= \max(\chi(G_1),\chi(G_2))$. If $G_i$ ($i \in 
\{1,2\}$) has a clique cutset, then we can recursively decompose 
$G_i$ in the same way.  This decomposition can be represented by 
a binary tree $T(G)$ whose root is $G$ and the two children of 
$G$ are $G_1$ and $G_2$, which are in turn the roots of subtrees 
representing the decompositions of $G_1$ and $G_2$. Each leaf of 
$T(G)$ corresponds to an induced subgraph of G that contains no 
clique cutset; we will call such graph an {\em atom}. 
Algorithmic aspects of the clique cutset decomposition are 
studied in Tarjan \cite{Tar1985} and Whiteside \cite{Whi1984}. In particular, the  
decomposition tree $T(G)$ can be constructed in $O(nm)$ time 
such that the total number atoms is at most 
$(n-1)$ \cite{Tar1985} (Here, as usual, $n$, resp., $m$, denotes the number of vertices, resp., edges, of the graph $G$).
This discussion can be summarized by the theorem below. 
\begin{Theorem}[\cite{Tar1985}, \cite{Whi1984}]\label{thm:clique-cutset-to-color}
Let $G$ be a graph. If every atom of $G$ can be colored in polynomial time, then $G$ can be colored in
polynomial time. \qed
\end{Theorem}
We will need the following theorem that illustrates the structure of ($4K_1, C_4, C_6, C_5$-twin)-free graphs.
\begin{Theorem}\label{thm:c5-twin}
	Let $G$ be a ($4K_1, C_4, C_6, C_5$-twin)-free graph. If $G$ contains a $C_5$, then one of the following holds: 
	\begin{description}
		\item [(i)] $G$ contains a clique cutset.
		\item [(ii)]$G$ contains a $C_7$.
		\item [(iii)]$G$ has bounded clique width.
		\item [(iv)]$G$ is the join of a (possibly empty) clique and a $C_5$. In this case, (iii) is also satisfied.
	\end{description}
	
\end{Theorem}
\noindent {\it Proof of Theorem~\ref{thm:c5-twin}}. Let $G$ be a ($4K_1, C_4, C_6, C_5-twin$)-free graph and suppose $G$ contains a $C_5$. Assume that $G$ contains no clique cutset and no $C_7$, for otherwise we are done. Define the sets $F_i, T_i, X_i, W, R$ as above. Note that $X_i = \emptyset$ for all $i$ because $G$ contains no $C_5$-twin. By Observation~\ref{obs:one-Fi}, at most one set $F_i$ can be non-empty. We will assume that this one set, it it exists, is $F_1$. Define $T = T_1 \cup \ldots \cup T_5$. 

We will show that
\begin{equation}\label{eq:R-is-empty}
\begin{minipage}{0.9\linewidth}
\begin{center}
$R = \emptyset$.
\end{center}
\end{minipage}
\end{equation}
Suppose $R \not= \emptyset$. Note that $W \cup T \cup F_1$ is a cutset, separating $R$ from the $C_5$. Let $C$ be a minimal ($R, C_5$)-separator of $G$ that is contained in $W \cup T \cup F_1$. By assumption, $C$ is not a clique. Consider two non-adjacent vertices $a,b$ in $C$. By the minimality of $C$, there is a chordless path $P$ with endpoints being $a,b$, and interior vertices  belonging to $R$. Since $R$ is a clique, $P$ has at most three edges.

Suppose first that $P$ has three edges. Enumerate the vertices of $P$ as $x, r_1, r_2, y$ with $r_i \in R$. Since $r_2$ is not adjacent to $a$, by Observation~\ref{obs:r-join-Fi-union-Ti}, we have $a \in W$. Similarily, we have $b \in W$. But by Observation~\ref{obs:r-clique}, $ab$ is an edge, a contradiction. 

So $P$ has two edges.  Enumerate the vertices of $P$ as $x, r, y$ with $r \in R$. Note that both $a,b$ have neighbors in the $C_5$. No vertex $c \in C_5$ can be adjacent to both $a,b$, for otherwise, there is a $C_4$ ($c,a,r,b$). So we have $a,b \in F_1 \cup T$, in particular, $a,b \notin W$.  Since $ab$ is not an edge, by Observation~\ref{obs:r-clique}, either $a$ or $b$, or both, belongs to some $T_s$. We may assume $b \in T_s$, that is, $b$ is a 2-vertex.

Let $i$ be a vertex in the $C_5$ that is adjacent to $a$. Let $j$ be a the vertex in $C_5$ that is adjacent to $b$ and is closest to $i$ in the $C_5$. Then $P' = (a, i, i+1, \ldots, j$) is an induced path. If $P'$ has length at least four, the $P \cup P'$ induces a chordless cycle of length at least six, a contradiction. So we know $j = i+1$. Since $b \in T_s$ and $bi$ is an edge, we know $b (i+3), b(i-1) \notin E$. It follows that $b (i+2) \in E$. The vertex $a$ may or may not be adjacent to $i-1$. If $a (i-1) \in E$, let   $P''$ be the chordless path $b, i+2, i+3, i-1, a$; otherwise, let $P''$ be the chordless path $b, i+2, i+3, i-1, i , a$. Then $P''$ and $r$ together induces a $C_6$ or $C_7$, a contradiction. We have establised (\ref{eq:R-is-empty}).

Next, we claim that
\begin{equation}\label{eq:F-is-not-empty}
\begin{minipage}{0.9\linewidth}
\begin{center}
$F \not= \emptyset$.
\end{center}
\end{minipage}
\end{equation}
Suppose $F = \emptyset$. If $T = \emptyset$, then $G$ is the join of $W$ and the $C_5$, and we are done. (Note that in this case $G$ has clique width three). We may assume some $T_i$ is not empty. By Observation~\ref{obs:Ti-cojoin-Tj}, the vertices in $T_i$ have no neighbors in $T_j$ with $i \not= j$. So $W \cup \{i, i+1\}$ is a clique cutset separating $T_i$ from $\{i+2, i+3, i+4\}$, a contradiction. We have established (\ref{eq:F-is-not-empty}). 

Now, we may assume $F = F_1$ is not empty. Suppose $T_2 \not= \emptyset$. By Observations~\ref{obs:Ti-cojoin-Tj} and \ref{obs:F-cojoin-Ti}, $T_2$'s neighbors belong to $W \cup \{2,3\}$. But then  $W \cup \{2,3\}$ is a clique cutset of $G$, a contradiction. So we have $T_2 = \emptyset$, and by symmetry, $T_4 = \emptyset$.

Suppose that $T_3 \not= \emptyset$. If $|T_3| \geq 2$, then there is a $C_5$-twin ($f, 1, 2, 3, t, t'$) for any $f \in F_1$ and $t,t' \in T_3$ (by Observation~\ref{obs:F-join-Ti}, $f$ is adjacent to $t,t'$). So we have $|T_3| = 1$. If $T_1 \not= \emptyset$, then there is a $C_5$-twin ($f, t_3, 3,2,t_1,1$) for any $f \in F_1$, $t_1 \in T_1$, and $t_3 \in T_3$. So we have $T_1 = \emptyset$, and by symmetry, $T_5 = \emptyset$. Consider the graph $G'$ obtained from $G$ by removing the six vertices of $C_5 \cup T_3$. The partition $W, F_1$ is a near-uniform partition of $G'$. By Theorem~\ref{thm:cliquewidth}, $G'$ has bounded clique width. By Observation~\ref{obs:folklore}, $G$ has bounded clique width. 

So, we may assume that $T_3 = \emptyset$. If $T_1 = \emptyset$, then $\{1,5\} \cup W$ is a clique cutset separation $F_1 \cup T_5$ from $\{2,3,4\}$, a contradictionl. So we have $T_1 \not= \emptyset$, and by symmetry, $T_5 \not= \emptyset$. Consider vertices $t_5 \in T_5, f \in F_1, t_1 \in T_1$. By Observations~\ref{obs:Ti-cojoin-Tj} and \ref{obs:F-join-Ti}, we have $f t_5, f t_1 \in E$, and $t_1 t_5 \notin E$. Now, there is a $C_7$ ($t_5, f, t_1,2,3,4,5$) and so (ii) holds. \qed

We are now in position to prove Theorem~\ref{thm:c5-twin-color}.

\noindent {\it Proof of Thereom~\ref{thm:c5-twin-color}}. 	Let $G$ be a ($4K_1, C_4, C_6, C_5$-twin)-free graph. We may assume that $G$ is not perfect, for otherwise, we may use the algorithm of \cite{GroLov1984} to color $G$. Since $G$ is $C_4$-free, $G$ contains no anti-hole of length at least six. So $G$ must contain a $C_5$ or $C_7$. If $G$ contains a $C_7$, then we are done by Theorem~\ref{thm:main-coloring}. So, we may assume that $G$ contains a $C_5$, but no $C_7$. By Theorem~\ref{thm:clique-cutset-to-color}, we only need to show that every atom of $G$ can be colored in polynomial time. Let $A$ be an atom of $G$ (an induced subgraph with no clique cutset). By Theorem~\ref{thm:c5-twin}, $A$ has bounded clique width, thus it can be colored in poylynomial time by Theorem~\ref{thm:Rao2007}. \qed

\section{Conclusions}\label{sec:conclusion}
In this paper, we studied the complexity of VERTEX COLORING for ($4K_1, C_4, C_6$)-free graphs. We showed the problem admits a polynomial time algorithm when the graph in our class has a $C_7$. We have not solved the problem when the graph contains a $C_5$. We leave this as an open problem. In addition, we designed a polynomial time algorithm for VERTEX COLORING for  ($4K_1, C_4, C_6$, $C_5$-twin)-free graphs.
We note that the more general problem to color a $(4K_1, C_4$)-free graph in polynomial time is still open. 

\begin{center}
{\bf Acknowledgement}
\end{center}
This work was supported by the Canadian Tri-Council Research
Support Fund. The authors A.M.H. and C.T.H. were each supported by
individual NSERC Discovery Grants. Author T.P.M was supported by 
an NSERC Undergraduate Student Research Award (USRA).  This work
was done by authors D.J.F. and K.H. in partial fulfillment of the
course requirements for CP493: Directed Research Project I in the
Department of Physics and Computer Science at Wilfrid Laurier
University.

\end{document}